\documentclass[aps,twocolumn,amsmath,amssymb,floatfix,showpacs,
superscriptaddress]{revtex4}

\usepackage[dvips]{graphicx}
\usepackage{color}
\definecolor{gold}{rgb}{0.85,0.66,0}
\definecolor{dblue}{rgb}{0,0,0.4}

\begin{document}

\title{\textcolor{dblue}{Spin transport through a quantum network: 
Effects of Rashba spin-orbit interaction and Aharonov-Bohm flux}}

\author{Moumita Dey}

\affiliation{Theoretical Condensed Matter Physics Division, Saha
Institute of Nuclear Physics, Sector-I, Block-AF, Bidhannagar,
Kolkata-700 064, India}

\author{Santanu K. Maiti}

\email{santanu.maiti@saha.ac.in}

\affiliation{Theoretical Condensed Matter Physics Division, Saha
Institute of Nuclear Physics, Sector-I, Block-AF, Bidhannagar,
Kolkata-700 064, India}

\affiliation{Department of Physics, Narasinha Dutt College, 129
Belilious Road, Howrah-711 101, India}

\author {S. N. Karmakar}

\affiliation{Theoretical Condensed Matter Physics Division, Saha
Institute of Nuclear Physics, Sector-I, Block-AF, Bidhannagar,
Kolkata-700 064, India}

\begin{abstract}
We address spin dependent transport through an array of diamonds in
the presence of Rashba spin-orbit (SO) interaction where each diamond 
plaquette is penetrated by an Aharonov-Bohm (AB) flux $\phi$. The diamond 
chain is attached symmetrically to two semi-infinite one-dimensional 
non-magnetic metallic leads. We adopt a single particle tight-binding 
Hamiltonian to describe the system and study spin transport using 
Green's function formalism. After presenting an analytical method for 
the energy dispersion relation of an infinite diamond chain in the 
presence of Rashba SO interaction, we study numerically the 
conductance-energy characteristics together with the density of states 
of a finite sized diamond network. At the typical flux $\phi=\phi_0/2$, 
a delocalizing effect is observed in the presence of Rashba SO interaction, 
and, depending on the specific choices of SO interaction strength and AB 
flux the quantum network can be used as a spin filter. Our analysis may 
be inspiring in designing spintronic devices.
\end{abstract}


\maketitle

\section{Introduction}

In recent times spin transport in low dimensional systems has drawn much 
attention both from theoretical as well as experimental point of view, due to 
its promising applications in the field of `spintronics'~\cite{spintronics}. 
It is a newly developed sub-discipline in condensed matter physics, that 
deals with the idea of manipulating spin of the electrons in transport 
phenomena in addition to their charge, and holds future promises to 
integrate memory and logic into a single device. Since the discovery of 
Giant Magnetoresistance (GMR) effect~\cite{gmr} in Fe-Cr magnetic 
multilayers revolutionary advancement has taken place in data processing, 
device making and quantum computation techniques. Today generation of pure 
spin current is a major challenge to us for further development in quantum 
computation. A more or less usual way of realization~\cite{trend1,trend2} 
of spin filters is by using ferromagnetic leads or by external magnetic 
field. But in the first case, spin injection from ferromagnetic leads is 
difficult due to large resistivity mismatch and for the second one the 
difficulty is to confine a very strong magnetic field into a small region, 
like, a quantum dot (QD). Therefore, attention is now being paid for 
modeling of spin filters using the intrinsic properties~\cite{intrinsic1,
intrinsic2,intrinsic3,intrinsic4,intrinsic5,intrinsic6} of the mesoscopic 
systems such as spin-orbit interaction or voltage bias. Studies on Rashba 
spin orbit interaction~\cite{rashba1,rashba2,rashba3,rashba4}, which is 
present in asymmetric heterostructures has made a significant impact in 
semiconductor spintronics as far as the control of spin dynamics is 
concerned. It is generally important in narrow gap semiconductors and 
its strength can be tuned by electrostatic means, e.g., applying external 
gate voltages~\cite{gate1,gate2,gate3}. Rashba SO interaction induces spin 
flipping through a mechanism known as D'yakonov-Perel'~\cite{Dyakonov} 
mechanism, which is a slow spin scattering process in which spin 
precession takes place around the Rashba field during transmission. 

Over the last few years quantum networks are becoming prospective candidates 
for studying transport phenomena because of the manifestation of several 
interesting features, like, quantum interference, interplay of AB flux and 
network geometry on electron localization, spin-orbit interaction induced 
delocalization, effect of disorder, electron-electron interaction, etc.  
In 2000, Vidal {\em et al.} have shown Hubbard interaction can destroy the 
localization induced by magnetic field in a diamond network~\cite{vidal1}. 
In some other works~\cite{vidal2,vidal3} they studied the general formalism 
to obtain conductance of any quantum networks and the effect of disorder 
and interaction. Latter in 2002, they considered Josephson-junction chain 
of diamonds in a magnetic field to show a local $Z_2$ symmetry at half 
flux-quantum~\cite{vidal4}.
It may be interesting to study the effects of Rashba SO coupling and AB 
flux in such quantum networks. Depending on their topology these geometries 
exhibit various striking spectral properties, and the interplay between AB 
flux and Rashba SO strength can also be explored. In 2005, Bercioux {\em et 
al.}~\cite{bercioux} considered the effect of AB flux and Rashba SO 
interaction on the energy averaged conductances of a finite sized diamond 
chain. They observed that in such a network spin-orbit interaction or AB 
flux can induce complete localization, while the presence of both of them 
can lead to the effect of weak anti-localization. The possibility to use 
such a diamond network 
as a spin filter was explored by Aharony {\em et al.} in 2008~\cite{aharony}. 
In 2009, there was another work by Chakrabarti {\em et al.}~\cite{sil}, 
where they have shown how such a diamond network can be implemented as a 
p-type or n-type semiconductor depending on the suitable choice of the 
on-site potentials of the atoms at the vertices of the network and the 
strength of magnetic flux penetrating each diamond plaquette. But the 
effect of spin-orbit interaction was not considered. Several other 
interesting theoretical works have been done considering this kind of
geometry. Gulacsi {\em et al.}~\cite{gul1,gul2} in 2007 shown the exact 
ground state of diamond Hubbard chain in magnetic field exhibits a wide 
range of striking properties, those are tunable by magnetic flux, electron 
density, etc. Peeters {\em et al.} considered quantum rings in presence 
of Rashba SO interaction and magnetic field to obtain various features of 
magnetoconductance~\cite{peeters1,peeters2,peeters3}. In our present work, 
we wish to explore the spectral 
and transport properties of a diamond network in the presence of both AB 
flux and Rashba SO interaction. We calculate spin conserved and spin flip 
conductances using single-particle Green's function formalism~\cite{green1} 
within a tight-binding framework for a finite sized diamond chain, which is 
compatible with the analytical dispersion relation obtained by 
renormalization group method for an infinite diamond network. Analysis of 
the spin-dependent conductances, dispersion relation and the density of 
states (DOS) provides an insight about the effect of Rashba SO interaction 
and AB flux on the localization behavior of the electrons. Finally, we show 
that, for some specific choices of the external parameters this finite sized 
diamond network can achieve a high degree of spin polarization. 

Our organization of the paper is as follows. Following a brief introduction 
(Section I), in Section II, we present the model and the theoretical 
formulation. Section III is on our work comprising an analytical form for 
the energy dispersion relation for an infinite diamond network, the 
numerical calculations of two-terminal conductance, DOS, discussion on
delocalizing effect in presence of SO interaction and demonstration of spin 
filtering action for a finite sized diamond array. At the end, the summary 
of our work will be available in Section IV.

\section{Model and theoretical formulation}

\begin{figure*}[ht]
{\centering \resizebox*{14cm}{4.5cm}{\includegraphics{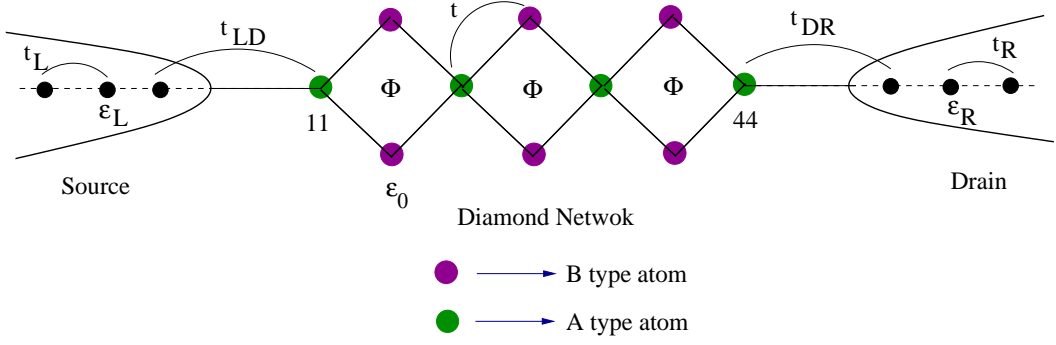}}\par}
\caption{(Color online). A finite sized diamond network (central region) 
connected to two semi-infinite one-dimensional non-magnetic metallic 
leads, viz, source and drain. The diamond network is composed of two 
types of atoms labeled by filled green and blue circles, where each 
diamond plaquette is penetrated by an AB flux $\phi$.}
\label{device1}
\end{figure*}

At the beginning of our theoretical formulation we start by describing 
the geometry of the quasi one-dimensional nanostructure through which 
spin transport properties are being investigated. In Fig.~\ref{device1} 
we illustrate schematically the quantum network, in which the square 
loops are connected at the vertices (termed as Diamond Network (DN) or 
Diamond Chain (DC)). The diamond array is connected symmetrically to two 
semi-infinite one-dimensional ($1$D) non-magnetic metallic leads, commonly 
known as source and drain which are characterized by the electrochemical 
potentials $\mu_1$ and $\mu_2$ under the non-equilibrium condition when 
a bias voltage is applied. 

The full Hamiltonian for the complete system, i.e., source-DN-drain can 
be written as,
\begin{equation}
H=H_{D}+H_L+H_R+H_{LD}+H_{DR}
\label{equ1}
\end{equation}
where, $H_{D}$ represents the Hamiltonian for the diamond network. 
$H_{L(R)}$ corresponds to the Hamiltonian for the left (right) lead, 
i.e., source (drain), and $H_{LD(DR)}$ is the Hamiltonian describing 
the chain-lead coupling.

\begin{figure}[ht]
{\centering \resizebox*{5.5cm}{3.5cm}{\includegraphics{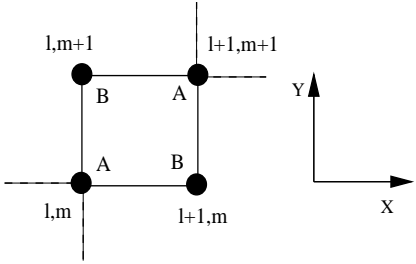}}\par}
\caption{(Color online). Index convention for representing the co-ordinates. 
$l$ and $m$ denote the co-ordinates along the $X$ and $Y$ directions, 
respectively.}
\label{config}
\end{figure}

We model the diamond network by the nearest-neighbor tight-binding 
Hamiltonian which in Wannier basis can be written as,
\begin{eqnarray}
H_{D} & = & \sum_{l,m} \mbox{\boldmath $c_{l,m}^{\dagger} \epsilon_0 
c_{l,m}$} + \sum_{l,m} \left(\mbox{\boldmath $c_{l,m}^{\dagger} t$} 
e^{i \alpha} \mbox {\boldmath $c_{l+1,m}$} + h.c. \right) \nonumber\\ 
& & + \sum_{l,m} \left(\mbox{\boldmath $c_{l,m}^{\dagger} t$} e^{i \alpha} 
\mbox{\boldmath $c_{l,m+1}$} + h.c. \right) \nonumber\\ 
& & + \sum_{l,m} \left(\mbox{\boldmath $c_{l,m}^{\dagger} 
(i \sigma_y)~t_{so}$} ~e^{i \alpha} \mbox{\boldmath $c_{l+1,m}$} + 
h.c. \right) \nonumber\\
& & - \sum_{l,m} \left(\mbox{\boldmath $c_{l,m}^{\dagger} 
(i \sigma_x)~t_{so}$}~e^{i \alpha} \mbox{\boldmath $c_{l,m+1}$} + h.c. 
\right)
\label{equ2}
\end{eqnarray}
where, \\
\mbox{\boldmath $c_{l,m}^{\dagger}$}=$\left(\begin{array}{cc}
c_{l,m \uparrow}^{\dagger} & c_{l,m \downarrow}^{\dagger} 
\end{array}\right);$
\mbox{\boldmath $c_{l,m}$}=$\left(\begin{array}{c}
c_{l,m \uparrow} \\
c_{l,m \downarrow}\end{array}\right);$
\mbox{\boldmath $\epsilon_0$}=$\left(\begin{array}{cc}
\epsilon_0 & 0 \\
0 & \epsilon_0 \end{array}\right);$ 
\mbox{\boldmath $t$}=$t\left(\begin{array}{cc}
1 & 0 \\
0 & 1 \end{array}\right);$ 
\mbox{\boldmath $t_{so}$}=$\left(\begin{array}{cc}
t_{so} & 0 \\
0 & t_{so} \end{array}\right)$. \\
~\\
\noindent
Here $\epsilon_0$ is the site energy of each atomic site of the diamond 
chain. For $A$ type of atoms $\epsilon_0=\epsilon_A$, while for $B$ type 
of atoms we call $\epsilon_0$ as $\epsilon_B$ (see Fig.~\ref{device1}).
The second and third terms represent the electron hopping along $X$ 
and $Y$ directions, respectively, where $t$ is the nearest-neighbor 
hopping strength and $\alpha=\frac{2\pi\phi}{4\phi_0}$ is the phase factor
due to the magnetic flux $\phi$ threaded by each diamond plaquette. Here 
we use double indexing to describe the location of lattice sites in the 
diamond network, as illustrated in Fig.~\ref{config} for a single plaquette. 
The fourth and fifth terms are associated with the spin dependent Rashba 
interaction, where $t_{so}$ is the isotropic nearest-neighbor transfer 
integral that measures the strength of Rashba SO coupling.

Similarly, the Hamiltonian $H_{L(R)}$ for the two leads can be written as,
\begin{equation}
H_{L(R)}=\sum_i \mbox{\boldmath $c_i^{\dagger} \epsilon_{L(R)} c_i$} + 
\sum_i \left(\mbox{\boldmath $c_i^{\dagger} t_{L(R)} c_{i+1}$} + h.c.\right).
\label{equ3}
\end{equation}
Here also, \\
~\\
\mbox{\boldmath $\epsilon_{L(R)}$}=$\left(\begin{array}{cc}
\epsilon_{L(R)} & 0 \\
0 & \epsilon_{L(R)} \end{array}\right);$~~~
\mbox{\boldmath $t_{L(R)}$}=$\left(\begin{array}{cc}
t_{L(R)} & 0 \\
0 & t_{L(R)} \end{array}\right)$ \\
~\\
\noindent
where, $\epsilon_{L(R)}$ is the site energy and $t_{L(R)}$ is the 
hopping strength between the nearest-neighbor sites in the left (right) 
lead.  

The diamond chain-to-lead coupling Hamiltonian is described by,
\begin{equation}
H_{LD(DR)}= \left(\mbox{\boldmath $c_{0(NN)}^{\dagger} t_{LD(DR)} 
c_{11(N+1)}$} + h.c.\right)
\label{equ4}
\end{equation}
where, $t_{LD(DR)}$ being the chain-lead coupling strength. 

In order to calculate spin dependent transmission probabilities through 
the quantum network, we use single particle Green's 
function formalism. Within the regime of coherent transport and in absence
of Coulomb interaction this technique is well applied.

The single particle Green's function representing the full system for 
an electron with energy $E$ is defined as,
\begin{equation}
G=(E-H+i\eta)^{-1}
\label{equ5}
\end{equation}
where $\eta \rightarrow 0^+$.

The matrix representation for the Hamiltonian can be expressed as 
\begin{equation}
\mbox{\boldmath $H$}=\left(\begin{array}{ccc}
\mbox{\boldmath $H_{L}$} & \mbox{\boldmath $H_{LD}$} & 0 \\
\mbox{\boldmath $H_{LD}^\dag$} & \mbox{\boldmath $H_{D}$} & 
\mbox{\boldmath $H_{DR}$}\\
0 & \mbox{\boldmath $H_{DR}^\dag$} & \mbox{\boldmath $H_{R}$}\\
\end{array} \right) 
\label{equ7}
\end{equation}
where, \mbox{\boldmath ${H_L}$}, \mbox{\boldmath ${H_D}$} and 
\mbox{\boldmath ${H_R}$} are the Hamiltonian matrices for the left lead, 
diamond network and right lead, respectively. \mbox{\boldmath ${H_{LD}}$}
and \mbox{\boldmath ${H_{DR}}$} are the coupling matrices between diamond 
network and the leads. Since there is no direct coupling between the leads 
themselves, the corner elements of \mbox{\boldmath $H$} are null matrices.
A similar definition holds true for the Green's function matrix 
\mbox{\boldmath $G$} as well.
\begin{equation}
\mbox{\boldmath $G$}=\left(\begin{array}{ccc}
\mbox{\boldmath $G_{L}$} & \mbox{\boldmath $G_{LD}$} & 0 \\
\mbox{\boldmath $G_{DL}$} & \mbox{\boldmath $G_{D}$} & 
\mbox{\boldmath $G_{DR}$}\\
0 & \mbox{\boldmath $G_{RD}$} & \mbox{\boldmath $G_{R}$}\\
\end{array} \right) 
\label{equ8}
\end{equation}
The problem of finding \mbox{\boldmath $G$} in the full Hilbert space
of \mbox{\boldmath $H$} can be mapped exactly to a Green's function
\mbox{\boldmath $G_D^{eff}$} corresponding to an effective Hamiltonian
in the reduced Hilbert space of diamond network and we have 
\begin{equation}
\mbox{\boldmath ${\mathcal G}$=$G_D^{eff}$}=\left(\mbox{\boldmath $E- H_D - 
\Sigma_L - \Sigma_R$}\right)^{-1}
\label{equ9}
\end{equation}
where, \mbox{\boldmath $\Sigma_L$} and  \mbox{\boldmath $\Sigma_R$} 
represent the contact self-energies introduced to incorporate the effects 
of semi-infinite leads coupled to the system. The self-energies are 
expressed by the relations,
\begin{eqnarray}
\mbox{\boldmath $\Sigma_L$} & = & \mbox{\boldmath $H_{LD}^{\dag} G_{L} 
H_{LD}$} \nonumber \\
\mbox{\boldmath $\Sigma_R$} & = &  \mbox{\boldmath $H_{DR}^{\dag} G_{R} 
H_{DR}$.}
\end{eqnarray}
Thus, the form of self-energies are independent of the nano-structure 
itself through which transmission is studied and they completely describe 
the influence of the two leads attached to the system. Now, the transmission 
probability $(T_{\sigma \sigma^{\prime}})$ of an electron with energy $E$ 
is related to the Green's function as,
\begin{eqnarray}
T_{\sigma \sigma^{\prime}} & = & \Gamma^{1}_{L(\sigma \sigma)} 
{\mathcal G}^{1N}_{r (\sigma \sigma^{\prime})} 
{\mathcal G}^{N1}_{a (\sigma^{\prime} \sigma)}
\Gamma^{N}_{R (\sigma^{\prime} \sigma^{\prime})} \nonumber \\
& = & \Gamma^{1}_{L (\sigma \sigma)} |{\mathcal G}^{1N}_{ (\sigma 
\sigma^{\prime})}|^2 \Gamma^{N}_{R (\sigma^{\prime} \sigma^{\prime})}
\label{equ11}
\end{eqnarray}
where,
$\Gamma^{1}_{L (\sigma \sigma)}$ = $\langle 11 \sigma| {\bf \Gamma_L} | 11 
\sigma \rangle $,
$\Gamma^{N}_{R (\sigma^{\prime} \sigma^{\prime})}$ = $\langle NN 
\sigma^{\prime}| {\bf \Gamma_R} |NN \sigma^{\prime} \rangle $ 
and ${\mathcal G}^{1 N}_{\sigma \sigma^{\prime}} =
\langle 11 \sigma| \mbox {\boldmath ${\mathcal G}$} |NN \sigma^{\prime} 
\rangle $.
Here, ${\mathcal G}_r$ and  ${\mathcal G}_{a}$ are the retarded and 
advanced single particle Green's functions for an 
electron with energy $E$. $\bf{\Gamma_{L}}$ and  $\bf{{\Gamma_{R}}}$ are 
the coupling matrices, representing the coupling of the quantum network 
to the left and right leads, respectively, and they are defined by the 
relation,
\begin{equation}
\mbox{\boldmath $\Gamma_{L(R)}$}=i \mbox{\boldmath $\left[\Sigma^r_{L(R)} - 
\Sigma^{a}_{L(R)}\right]$} 
\label{equ12}
\end{equation}
Here, \mbox{\boldmath${\Sigma^r_{L(R)}}$} and \mbox{\boldmath
${\Sigma^a_{L(R)}}$} are the retarded 
and advanced self-energies, respectively, and they are conjugate to each 
other. It is shown in literature by Datta {\em et al.}~\cite{green1} that 
the self-energy can be expressed as a linear combination of a real and an 
imaginary part in the form,
\begin{equation}
\mbox{\boldmath ${\Sigma^r_{L(R)}}$} = \mbox{\boldmath $\Lambda_{L(R)}$} - 
i \mbox{\boldmath $\Delta_{L(R)}$}
\label{equ13}
\end{equation}
The real part of self-energy describes the shift of the energy levels
and the imaginary part corresponds to the broadening of the levels. The 
finite imaginary part appears due to incorporation of the semi-infinite 
leads having continuous energy spectrum. Therefore, the coupling matrices 
can easily be obtained from the self-energy expression and is expressed as,
\begin{equation}
\mbox{\boldmath $\Gamma_{L(R)}$}=-2~{\mbox {Im}} 
(\mbox{\boldmath $\Sigma_{L(R)}$})
\label{equ14}
\end{equation}
Considering linear transport regime, conductance $(g_\sigma)$ is obtained
using two-terminal Landauer conductance formula,
\begin{equation}
g_{\sigma \sigma^{\prime}}=\frac{e^2}{h}T_{\sigma \sigma^{\prime}}
\label{equ15}
\end{equation}
Throughout our study we choose $c=e=h=1$ for simplicity.

\section{Numerical results and discussion}

In this section we study spin dependent transport through a diamond 
chain in presence of Rashba spin orbit interaction and magnetic flux and 
investigate the interplay between them. An array of diamonds is a 
bipartite structure 
\begin{figure*}[ht]
{\centering \resizebox*{5cm}{12cm}{\includegraphics{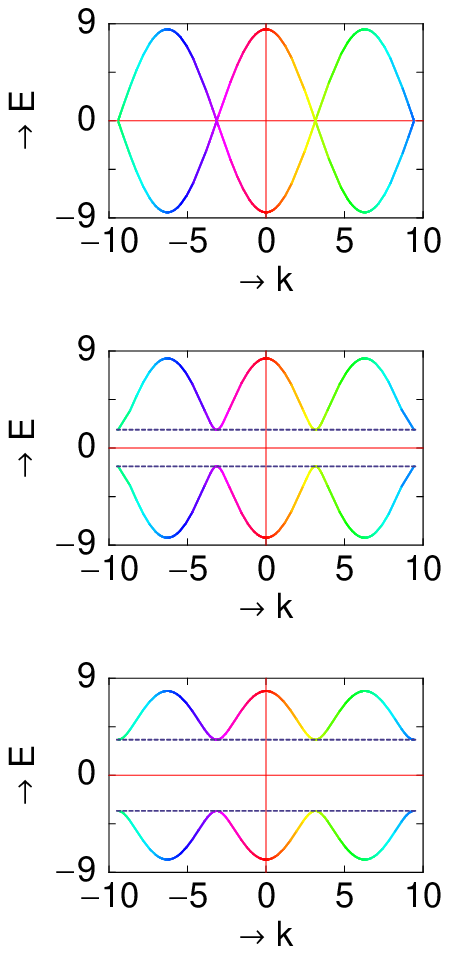}}
\resizebox*{5cm}{12cm}{\includegraphics{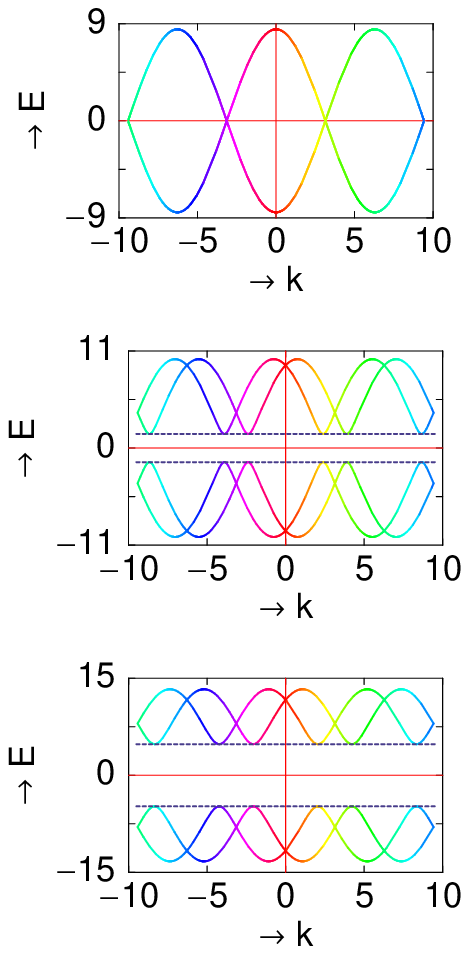}}
\resizebox*{5cm}{12cm}{\includegraphics{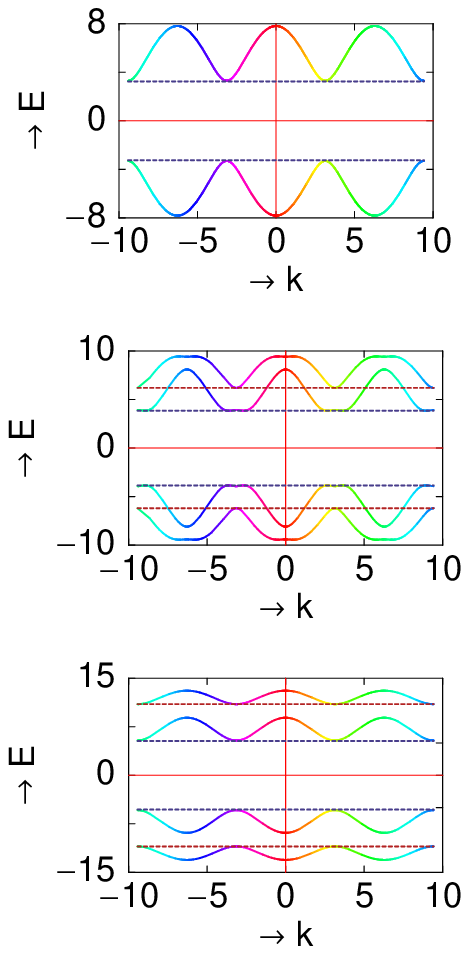}}\par}
\caption{(Color online). Energy dispersion ($E$-$k$) curves for an 
infinite diamond chain with $\epsilon_A=\epsilon_B=0$. The upper, 
middle and lower spectra in the $1$st column correspond to $\phi=0$, 
$0.2$ and $0.4$, respectively, when $t_{so}=0$. In the $2$nd column 
three different spectra from the top represent the results for $t_{so}=0$, 
$2$ and $4$, respectively, when $\phi$ is set to $0$. Finally, the three 
different figures in the last column refer to the results for the identical 
values of $t_{so}$ considered in the $2$nd column when $\phi$ is fixed at 
$0.4$.}
\label{dispersion}
\end{figure*}
with lattice sites having different co-ordination numbers. Electron 
localization plays a significant role even in the absence of disorder 
in this kind of geometry due to quantum interference effect. First we 
obtain analytically the dispersion relation for an infinite diamond chain 
in the presence of magnetic flux and Rashba interaction. Next, we simulate
numerically various features of spin transport using a finite size diamond 
chain. Before analyzing the results first we specify the values of the 
parameters those are used in the numerical simulations. We consider that 
the two non-magnetic side-attached leads are made up of identical materials. 
The on-site energies in the two leads ($\epsilon_{L(R)}$) are set to $0$. 
Hopping strength between the sites in the leads is chosen as $t_{L(R)}=4$, 
whereas in the diamond chain it is set as $t=3$. The Rashba strength 
($t_{so}$) is chosen to be uniform along $X$ and $Y$ directions and throughout 
the calculation its magnitude is considered as comparable to $t$. Energy 
scale is fixed in unit of $t$. Throughout the analysis we present all the 
results considering the chain-to-electrode coupling strength as 
$t_{LD} = t_{DR} = 2.5$.

\subsection{Energy dispersion relation in presence of Rashba SO interaction
and magnetic flux}

The energy dispersion relation for an infinite diamond chain clearly depicts
several significant features of this kind of topology. In order to study
the $E$-$k$ relation theoretically, first we map the quasi one-dimensional
diamond network into a linear chain with modified site energy and hopping 
strength.                                                                

We start with the Schrodinger equation which can be cast in the form of a
difference equation. For an arbitrary site ($n,p$), where $n$ and $p$ denote 
the indexing along $X$ and $Y$ directions, respectively, the difference 
equation can be expressed as
\begin{eqnarray}
\mbox{\boldmath ${(E-\epsilon)\psi_{np}}$} & = & e^{\mp i \alpha}
\mbox{\boldmath $t_{x_+} \psi_{n+1,p}$} + e^{\mp i \alpha}
\mbox{\boldmath $t_{x_-} \psi_{n-1,p}$} + \nonumber\\ 
& & e^{\pm i \alpha} \mbox{\boldmath $t_{y_+} \psi_{n,p+1}$} 
+ e^{\pm i \alpha} \mbox{\boldmath $t_{y_-} \psi_{n,p-1}$}
\label{d1}
\end{eqnarray}
where,\\
\mbox {\boldmath $t_{x_+}$}=$\left(\begin{array}{cc}
t & t_{so} \\
-t_{so} & t \end{array}\right);$ 
\mbox {\boldmath $t_{x_-}$}=$\left(\begin{array}{cc}
t & -t_{so} \\
t_{so} & t \end{array}\right);$ \\ \\
\mbox {\boldmath $t_{y_+}$}=$\left(\begin{array}{cc}
t & i t_{so} \\
i t_{so} & t \end{array}\right);$ 
\mbox {\boldmath $t_{y_-}$}=$\left(\begin{array}{cc}
t & - i t_{so} \\
- i t_{so} & t \end{array}\right);$ \\ \\
\mbox {\boldmath $E$}=$\left(\begin{array}{cc}
E & 0 \\
0 & E \end{array}\right)$;
\mbox {\boldmath $\epsilon$}=$\left(\begin{array}{cc}
\epsilon_{0} & 0 \\
0 & \epsilon_{0} \end{array}\right)$;
and
\mbox {\boldmath $\psi_{np}$}=$\left(\begin{array}{c}
\psi_{np,\uparrow} \\
\psi_{np,\downarrow} \end{array}\right)$ \\ \\
\noindent
$\phi$ being the AB flux enclosed by each diamond plaquette, 
$\psi_{np\sigma}$ being the wave function amplitude at the np-th site 
with spin $\sigma$. $\phi_0=ch/e$, the elementary flux-quantum. 

We begin the decimation technique by writing down the difference equations
at the sites containing A and B type atoms (see Fig.~\ref{config}). The 
equations are given below
\begin{eqnarray}
\mbox {\boldmath $(E-\epsilon_B)\psi_{12}$} & = & e^{i \alpha} 
\mbox {\boldmath $t_{x_+}\psi_{22}$} + e^{-i \alpha} 
\mbox {\boldmath $t_{y_-} \psi_{11}$} \nonumber\\
\mbox {\boldmath $(E-\epsilon_B)\psi_{21}$} & = & e^{i \alpha} 
\mbox {\boldmath $t_{x_-}\psi_{11}$} + e^{-i \alpha} 
\mbox {\boldmath $t_{y_+} \psi_{22}$} \nonumber\\
\mbox {\boldmath $(E-\epsilon_B)\psi_{23}$} & = & e^{i \alpha} 
\mbox {\boldmath $t_{x_+}\psi_{33}$} + e^{-i \alpha} 
\mbox {\boldmath $t_{y_-} \psi_{22}$} \nonumber\\
\mbox {\boldmath $(E-\epsilon_B)\psi_{32}$} & = & e^{i \alpha} 
\mbox {\boldmath $t_{x_-}\psi_{22}$} + e^{-i \alpha} 
\mbox {\boldmath $t_{y_+} \psi_{33}$}
\label{d2}
\end{eqnarray}
and,
\begin{eqnarray}
\mbox {\boldmath $(E-\epsilon_A)\psi_{22}$} & = & e^{-i \alpha} 
\mbox {\boldmath $t_{x_+}\psi_{32}$} + e^{-i \alpha}
\mbox {\boldmath $t_{x_-} \psi_{12}$} \nonumber\\
& & + e^{i \alpha} \mbox {\boldmath $t_{y_+}\psi_{23}$} + e^{i \alpha} 
\mbox {\boldmath $t_{y_-}\psi_{33}$} 
\label{d3}
\end{eqnarray}
Substituting \mbox {\boldmath $\psi_{32}$}, \mbox {\boldmath $\psi_{12}$}, 
\mbox {\boldmath $\psi_{23}$} and \mbox  {\boldmath $\psi_{33}$} 
from Eq.~(\ref{d2}) in Eq.~(\ref{d3}) we get,
\begin{equation}
\mbox {\boldmath $(E-\epsilon^{\prime}) \psi_{22}$} = \mbox {\boldmath 
$t_b \psi_{11} + t_f \psi_{33}$}
\label{d4}
\end{equation}
This represents the difference equation for an infinite linear chain with 
modified site energy \mbox{\boldmath $\epsilon^{\prime}$} and the forward 
and backward hopping strengths ${\bf t_{f}}$ and ${\bf t_{b}}$, respectively.
These quantities are expressed as follows.
\begin{eqnarray}
\mbox {\boldmath ${\epsilon^{\prime}}$} & = & \mbox {\boldmath $\epsilon_A$} 
+ \mbox{\boldmath $t_{x_+}.(E-\epsilon_B)^{-1}.t_{x_-}$} \nonumber \\
 & & + \mbox{\boldmath $t_{x_-}.(E-\epsilon_B)^{-1}.t_{x_+}$} 
+ \mbox{\boldmath $t_{y_+}.(E-\epsilon_B)^{-1}.t_{y_-}$} \nonumber \\
 & & + \mbox{\boldmath $t_{y_-}.(E-\epsilon_B)^{-1}.t_{y_+}$} \nonumber\\
\mbox {\boldmath $t_b$} & = & e^{-2 i \alpha} \mbox {\boldmath 
$t_{x_-}.(E-\epsilon_B)^{-1}.t_{y_-}$} \nonumber \\
& & + e^{2 i \alpha} \mbox {\boldmath $t_{y_-}.(E-\epsilon_B)^{-1}.t_{x_-}$} 
\nonumber\\
\mbox {\boldmath $t_f$} & = & e^{-2 i \alpha} \mbox {\boldmath 
$t_{x_+}.(E-\epsilon_B)^{-1}.t_{y_+}$} \nonumber \\
& & + e^{2 i \alpha} \mbox {\boldmath $t_{y_+}.(E-\epsilon_B)^{-1}.t_{x_+}$} 
\nonumber\\
\label{d5}
\end{eqnarray}
As the translational invariance is preserved in this decimated infinite
linear chain, the solution will be of Bloch form and can be written as,
\begin{equation}
\mbox {\boldmath $\psi_{n}$} = \sum_k e^{i k n a} 
\left(\begin{array}{c}
\psi_{k,\uparrow} \\
\psi_{k,\downarrow} \end{array}\right)
\label{d6}
\end{equation}
\mbox {\boldmath $\psi_n$} being a short form of \mbox {\boldmath $\psi_{nn}$}.

Using this form of \mbox {\boldmath $\psi_n$}, the difference equation for 
an arbitrary site $n$ can be expressed as,
\begin{eqnarray} 
\sum_k \mbox {\boldmath $(E-\epsilon^{\prime})$} \left (\begin{array}{c}
\psi_{k,\uparrow} \\
\psi_{k,\downarrow} \end{array} \right) e^{i k n a} & = & 
\mbox {\boldmath $t_f$}
\sum_k \left (\begin{array}{c}
\psi_{k,\uparrow} \\
\psi_{k,\downarrow} \end{array} \right) e^{i k (n+1) a} \nonumber\\
& + & \mbox {\boldmath $t_b$}
\sum_k \left (\begin{array}{c}
\psi_{k,\uparrow} \\
\psi_{k,\downarrow} \end{array} \right) e^{i k (n-1) a} \nonumber \\
\label{d7}
\end{eqnarray}
For the non-trivial solution of Eq.~(\ref{d7}) we have the relation,
\begin{equation}
{\bf Det[M]} = 0
\label{d8}
\end{equation}
where, \mbox {\boldmath $M$}=$(\mbox {\boldmath $E$}- \mbox {\boldmath 
$\epsilon^{\prime}$}
-\mbox {\boldmath $t_f$} e^{ika} -\mbox {\boldmath $t_b$} e^{-ika})$ .

Expanding Eq.~(\ref{d8}) we obtain a $4$-th degree polynomial in $E$ and 
solving it we get the $E$-$k$ dispersion relation in terms of the parameters
$\phi$ and $t_{so}$. The solutions correspond to energy eigenstates which 
are linear combinations of up $(|k \uparrow \rangle)$ and down $(|k 
\downarrow \rangle)$ states.

Following the above analytical treatment, in Fig.~\ref{dispersion} we
show the $E$ versus $k$ dispersion curves for some typical parameter values 
of $\phi$ and $t_{so}$. The first column corresponds to the results for 
some specific values of AB flux $\phi$ in the absence of
Rashba SO coupling, i.e., $t_{so}=0$. It is observed that for zero magnetic 
flux the spectrum is degenerate and gapless, whereas a small non-zero flux 
$(\phi=0.2)$ opens a gap symmetrically around $E=0$ preserving the 
degeneracy. Here we choose $\epsilon_A=\epsilon_B=0$ and the gap appears 
symmetrically as long as $\phi$ is introduced, but the point is that for 
unequal values of $\epsilon_A$ and $\epsilon_B$ gap always appears even 
in the absence of $\phi$ (which is not shown in the figure). The width 
of the gap increases symmetrically with the rise in $\phi$. 
In the second column of Fig.~\ref{dispersion} we present the energy
dispersion curves for the three different values of Rashba SO coupling
strength keeping $\phi=0$, where the upper, middle and lower spectra
correspond to $t_{so}=0$, $2$ and $4$, respectively. The upper spectrum
is gapless and degenerate as described earlier. For non-zero values of 
$t_{so}$, the energy spectra get splitted vertically and all the degeneracies 
are removed except at the points $k=n\pi$, where $n=0$, $\pm 1$, $\pm 2$,
$\dots$. The gap becomes widened with the increase in Rashba strength as
clearly noticed from the middle and lower spectra. 
The above features seem to be more interesting when a non-zero magnetic 
flux is applied. In this case, each sub-band gets separated vertically 
as illustrated in the third column of Fig.~\ref{dispersion}. With a 
\begin{figure}[ht]
{\centering \resizebox*{8cm}{7cm}{\includegraphics{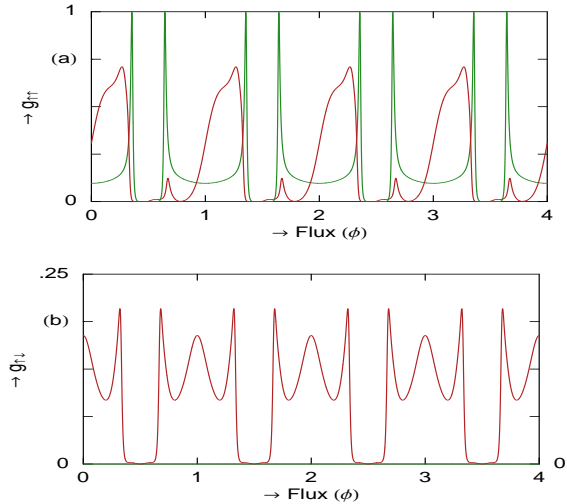}}\par}
\caption{(Color online). Variations of (a) $g_{\uparrow \uparrow}$ and
(b) $g_{\uparrow \downarrow}$ with AB flux $\phi$ for a diamond chain
considering $5$ plaquettes at the typical energy $E=5$. The green and
red curves correspond to $t_{so}=0$ and $2$, respectively. Here we set
$\epsilon_A=\epsilon_B=0$.} 
\label{flux}
\end{figure}
sufficiently high magnetic flux ($\phi=0.4$) and Rashba strength ($t_{so}=4$) 
two additional gaps occur in the dispersion spectrum along with the 
previous one, and these gaps can be {\em controlled externally by tuning 
the AB flux or the Rashba strength}. We will show that these results are
quite significant so far as the designing of nanoscale spintronic
devices are concerned.
                                   
\subsection{Variation of conductances with AB flux}

In Fig.~\ref{flux} we plot conductance-flux characteristics for a diamond
network both in the presence and absence of Rashba SO interaction. The
results are computed at a typical energy $E=5$ considering five diamond
plaquettes, where the green and red curves correspond to $t_{so}=0$ and $2$,
respectively. It is observed that in the absence of Rashba coupling spin flip
conductance $g_{\uparrow \downarrow}$ drops exactly to zero for the
entire range of $\phi$ (green curve in Fig.~\ref{flux}(b), coincides with
the $\phi$ axis), while spin conserved conductance $g_{\uparrow \uparrow}$ 
persists and it provides $\phi_0$ flux-quantum periodicity as a function 
of $\phi$. Interestingly we see that $g_{\uparrow \uparrow}$ completely 
vanishes at $\phi=\phi_0/2$ (see green curve of Fig.~\ref{flux}(a)) due to 
the complete destructive interference among the electronic waves passing 
through different arms of the plaquettes. On the other hand, in the
presence of Rashba SO interaction both $g_{\uparrow \uparrow}$ and 
$g_{\uparrow \downarrow}$ have values for wide ranges 
of $\phi$ and a significant change in their amplitudes takes place 
compared to the case where $t_{so}=0$. In the presence of the SO interaction,
the oscillatory character of the conductances is still preserved providing 
traditional $\phi_0$ flux-quantum periodicity. The important feature is 
that even for non-zero value of $t_{so}$, spin flip conductance disappears
at $\phi=\phi_0/2$. Since $g_{\downarrow \downarrow}$ and $g_{\downarrow 
\uparrow}$ exhibit exactly identical behavior to those mentioned for 
$g_{\uparrow \uparrow}$ and $g_{\uparrow \downarrow}$, respectively, we 
do not display these results explicitly.

The above numerical results can be justified from the following 
mathematical analysis. 

\begin{figure}[ht]
{\centering \resizebox*{7cm}{2.5cm}{\includegraphics{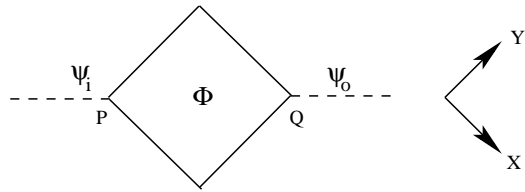}}\par}
\caption{(Color online). A single diamond plaquette threaded by an AB flux
$\phi$. $\psi_i$ and $\psi_o$ denote the incoming and outgoing waves,
respectively.}
\label{ab}
\end{figure}

To illustrate the behaviors of AB oscillation both in the presence
and absence of Rashba SO interaction, we consider an ideal $1$D square 
loop threaded by an AB flux $\phi$, as shown schematically in Fig.~\ref{ab}.
Two semi-infinite one-dimensional leads are connected at the vertices 
P and Q of the square loop. 
Rashba SO interaction is considered to be present only in the loop and
not in the leads. If \mbox{\boldmath $\psi_i$} and \mbox{\boldmath $\psi_o$} 
describe the incoming and outgoing wave functions at the respective vertices 
P and Q, then \mbox{\boldmath $\psi_o$} can be obtained considering only 
$1$st order tunneling processes as~\cite{xie}, 
\begin{equation}
\mbox{\boldmath $\psi_0$} = \frac{1}{2} \left(e^{-i\frac{\gamma}{2}} 
\mbox{\boldmath $R_x$}(\theta) \mbox{\boldmath $R_y$} (\theta) + 
e^{i\frac{\gamma}{2}} \mbox{\boldmath $R_y$} (\theta) 
\mbox {\boldmath $R_x$}(\theta)\right) \mbox{\boldmath $\psi_i$}
\label{abequ1}
\end{equation}
where, $\gamma=\frac{2\pi \phi}{\phi_0}$. {\boldmath $R_{\hat{r}}$}$(\theta)$ 
is the rotation operator defined by the relation,
\begin{equation}
\mbox{\boldmath $R_{\hat{r}}$}(\theta) = \mbox{\boldmath $I$} \cos 
\frac{\theta}{2} - i \hat{r}.\vec{\sigma} \sin \frac{\theta}{2}
\label{abequ2}
\end{equation}
where, $\theta=\frac{2m^*\alpha_R L}{\hbar^2}$ is the spin precession 
angle. $\alpha_R$ is the strength of Rashba SO interaction and $L$ 
represents the length of each side of the square loop. The wave functions
{\boldmath $\psi_i$} and {\boldmath $\psi_o$} used in Eq.~(\ref{abequ1}) 
are defined as follows. 
\begin{center}
{\boldmath $\psi_o$} $=\left(\begin{array}{c}
\psi_{o,\uparrow} \\
\psi_{o,\downarrow} \end{array}\right)$
and
{\boldmath $\psi_i$} $=\left(\begin{array}{c}
\psi_{i,\uparrow} \\
\psi_{i,\downarrow} \end{array}\right)$.
\end{center}
Following Eq.~(\ref{abequ2}) the matrices {\boldmath $R_x$}$(\theta)$ and
{\boldmath $R_y$}$(\theta)$ can be written as given below.
\begin{center}
{\boldmath $R_x$}$(\theta)=\left(\begin{array}{cc}
\cos\frac{\theta}{2} & -i \sin\frac{\theta}{2} \\
-i \sin\frac{\theta}{2} & \cos\frac{\theta}{2} \end{array}\right)$.
\end{center}
and
\begin{center}
{\boldmath $R_y$}$(\theta)=\left(\begin{array}{cc}
\cos\frac{\theta}{2} & - \sin\frac{\theta}{2} \\
 \sin\frac{\theta}{2} & \cos\frac{\theta}{2} \end{array}\right)$.
\end{center}
With these matrix forms we can express the wave functions 
$|\psi_{o,\uparrow}\rangle$ and $|\psi_{o,\downarrow}\rangle$ as linear 
combinations of $|\psi_{i,\uparrow}\rangle$ and $|\psi_{i,\downarrow}\rangle$ 
by expanding Eq.~(\ref{abequ1}) as, 
\begin{eqnarray}
|\psi_{o,\uparrow}\rangle & = & c_{\uparrow \uparrow} |\psi_{i,\uparrow}
\rangle + c_{\downarrow \uparrow}|\psi_{i,\downarrow}\rangle \nonumber \\
|\psi_{o,\downarrow}\rangle & = & c_{\uparrow \downarrow} |\psi_{i,\uparrow}
\rangle + c_{\downarrow \downarrow}|\psi_{i,\downarrow}\rangle
\label{abequ3}
\end{eqnarray}
where, the co-efficients $c_{\sigma \sigma^{\prime}}$ are functions of
$\theta$ and $\phi$. 

Now the probability of getting an up spin electron at the point Q, for the 
incidence of an electron with up spin at the point P, i.e., the spin 
conserved transmission probability $T_{\uparrow \uparrow}$ is proportional 
to $|\langle \psi_{i,\uparrow}|\psi_{o,\uparrow}\rangle|^2$, viz, 
$|c_{\uparrow \uparrow}|^2$. Similarly, the probability of getting a down
spin electron with up spin incidence, i.e., the spin flip transmission 
probability $T_{\uparrow \downarrow}$ is proportional to
$|\langle \psi_{i,\uparrow}|\psi_{o,\downarrow}\rangle|^2$, viz,
$|c_{\uparrow \downarrow}|^2$. After a few mathematical steps the 
quantities $|c_{\uparrow \uparrow}|^2$ and $|c_{\uparrow \downarrow}|^2$
are expressed as,
\begin{eqnarray}
|c_{\uparrow \uparrow}|^2 &=& \frac{1}{8} e^{-i\frac{2\pi \phi}{\phi_0}}
\left[1 + i\cos\theta + e^{i\frac{2\pi \phi}{\phi_0}} (i + \cos \theta) 
\right] \nonumber \\
 & & \times \left[\cos\theta -i + e^{i\frac{2\pi \phi}{\phi_0}} (1 -i 
\cos \theta) \right] 
\label{abequ4}
\end{eqnarray}
and
\begin{equation}
|c_{\uparrow \downarrow}|^2=\frac{1}{8} e^{-i\frac{2\pi \phi}{\phi_0}}
\left(1 + e^{i\frac{2\pi \phi}{\phi_0}} \right)^2 \sin^2\theta.
\label{abequ5}
\end{equation}
In the absence of Rashba SO interaction $\theta=0$ and the above two 
equations can be simplified as follows.
\begin{equation}
|c_{\uparrow \uparrow}|^2=\frac{1}{2} \left[1+\cos\left(\frac{2 \pi \phi}
{\phi_0} \right) \right]
\label{abequ6}
\end{equation} 
and
\begin{equation}
|c_{\uparrow \downarrow}|^2 = 0
\label{abequ7}
\end{equation} 
With the last four mathematical expressions 
(Eqs.~(\ref{abequ4})-(\ref{abequ7})) we can clearly justify the essential 
features those are presented in Fig.~\ref{flux}. In the absence of Rashba 
SO interaction, spin flip conductance vanishes for the entire range of 
$\phi$ (coincident green curve of Fig.~\ref{flux}(b) with $\phi$ axis) 
in accordance with Eq.~(\ref{abequ7}). On the other hand, a oscillatory 
character of up spin conductance with $\phi_0$ periodicity in the absence 
of $t_{so}$ (green curve of Fig.~\ref{flux}(a)) follows from 
Eq.~(\ref{abequ6}). 
The vanishing behavior of $g_{\uparrow \uparrow}$ at the typical flux 
$\phi=\phi_0/2$ is also justified from Eq.~(\ref{abequ6}). In the presence 
of SO interaction, both pure spin transmission and spin flip transmission 
get modified satisfying Eqs.~(\ref{abequ4}) and (\ref{abequ5}), respectively. 
For finite value of $t_{so}$, spin flip conductance always vanishes at 
$\phi=\phi_0/2$ obeying Eq.~(\ref{abequ5}).

\subsection{Conductance-energy characteristics}

Now we focus our attention on the conductance-energy characteristics of a 
finite sized diamond network for some specific values of AB flux $\phi$
and Rashba SO interaction strength $t_{so}$.

\begin{figure}[ht]
{\centering \resizebox*{9cm}{10cm}{\includegraphics{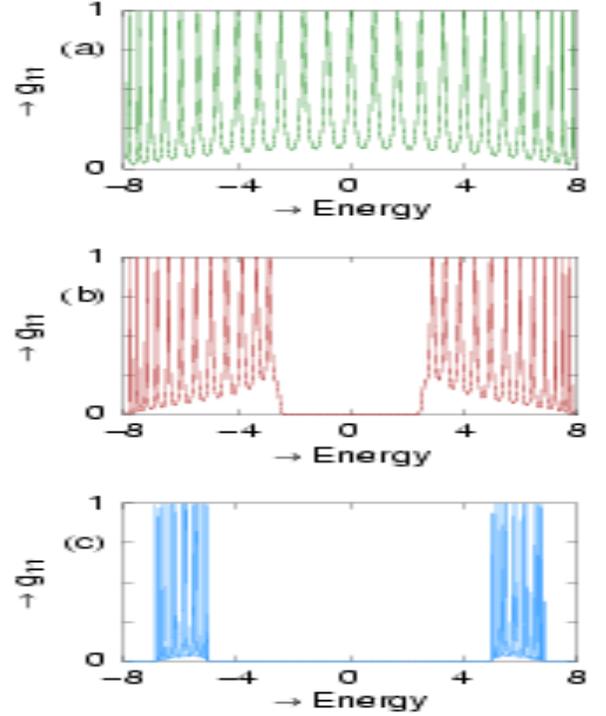}}\par}
\caption{(Color online). Conductance-energy ($g_{\uparrow \uparrow}$-$E$) 
characteristics in the absence of Rashba SO strength for a diamond network 
considering $15$ diamond plaquettes with $\epsilon_A=\epsilon_B=0$. (a), 
(b) and (c) correspond to $\phi=0$, $0.2$ and $0.4$, respectively.}
\label{cond1}
\end{figure}

In Fig.~\ref{cond1} we plot up spin conductances ($g_{\uparrow \uparrow}$)
as a function of injecting electron energy ($E$) for a diamond network
considering $15$ identical plaquettes in the absence of Rashba interaction. 
The top, middle and bottom spectra correspond to AB flux $\phi=0$, $0.2$
and $0.4$, respectively. When $\phi=0$, the spectrum is gapless
(Fig.~\ref{cond1}(a)). The presence of $\phi$ opens a gap and the width 
of the gap increases with the rise in $\phi$ as evident from 
\begin{figure*}[ht]
{\centering \resizebox*{8cm}{10cm}{\includegraphics{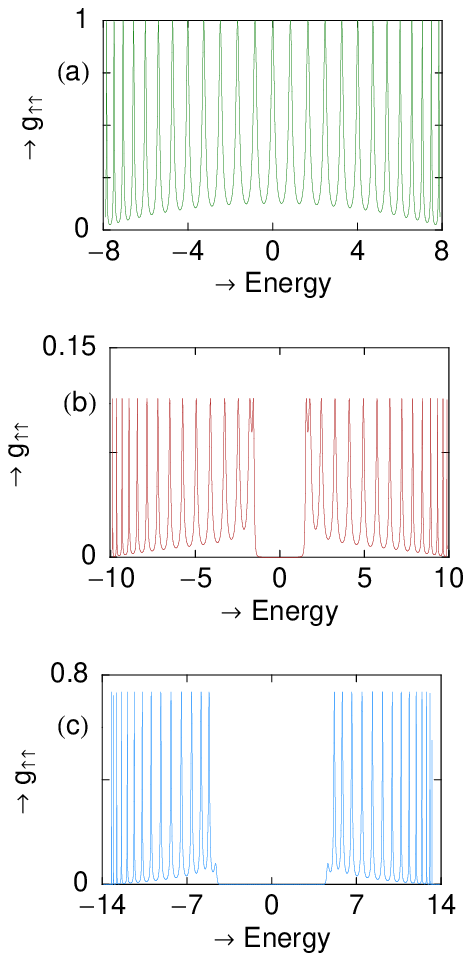}}
\resizebox*{8cm}{10cm}{\includegraphics{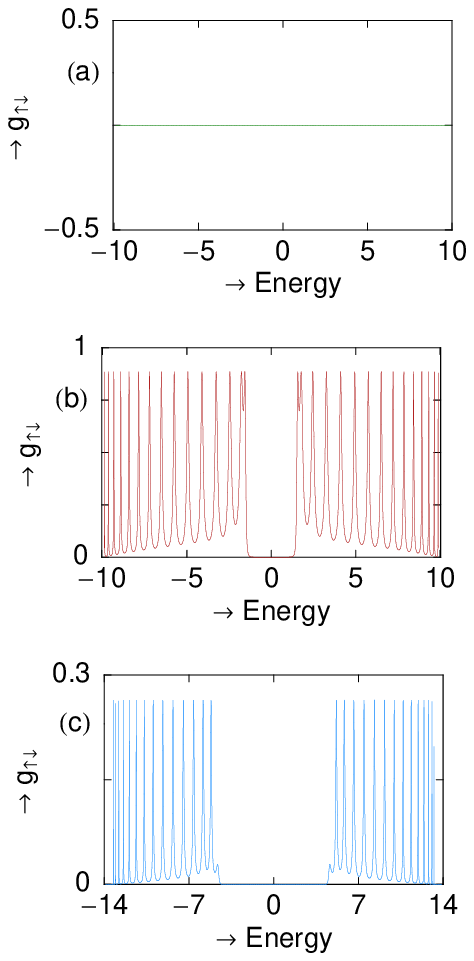}}\par}
\caption{(Color online). $g_{\uparrow \uparrow}$ and 
$g_{\uparrow \downarrow}$ as a function of energy $E$ for a diamond chain
with $15$ plaquettes considering $\epsilon_A=\epsilon_B=0$ in the absence 
of AB flux $\phi$. The $1$st, $2$nd and $3$rd rows represent the results 
when $t_{so}=0$, $2$ and $4$, respectively.}
\label{cond23}
\end{figure*}
\begin{figure}[ht]
{\centering \resizebox*{9.5cm}{7cm}{\includegraphics{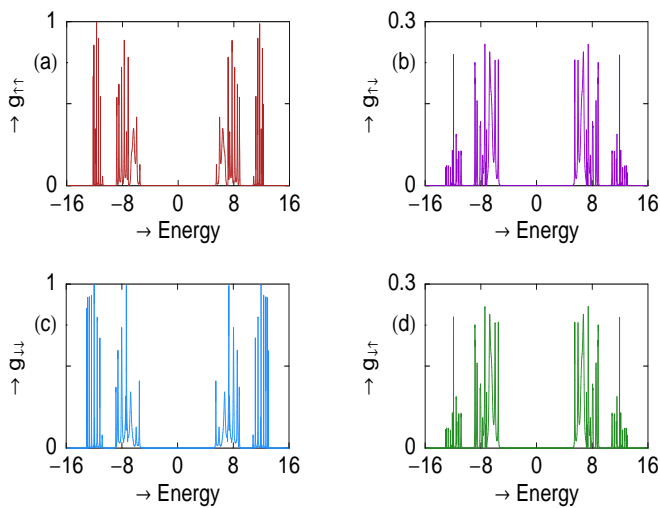}}\par}
\caption{(Color online). Spin conserved ($g_{\uparrow \uparrow}$, 
$g_{\downarrow \downarrow}$) and spin flip conductances 
($g_{\uparrow \downarrow}$, $g_{\downarrow \uparrow}$) as a function of
energy $E$ for a diamond chain with $15$ plaquettes when $\phi$ and 
$t_{so}$ are fixed to $0.4$ and $4$, respectively. The parameters 
$\epsilon_A$ and $\epsilon_B$ are set to $0$.}
\label{cond4}
\end{figure}
Figs.~\ref{cond1}(b) and (c). This gap is symmetric around the energy
$E=0$ with the choice $\epsilon_A=\epsilon_B=0$. On the other hand,
if the site energies $\epsilon_A$ and $\epsilon_B$ are unequal then a
gap in the conductance spectrum appears (not shown in the figure) even 
in the absence of magnetic flux both for an infinite as well as for a 
finite sized diamond array. This gap will be symmetric across $E=0$
provided $\epsilon_A$ and $\epsilon_B$ are identical, and the width 
of the gap increases symmetrically about the center of the gap with 
the enhancement in $\phi$. In this particular case we do not consider 
any Rashba interaction, and therefore, no spin flip transmission takes 
place.

In Fig.~\ref{cond23} we show the conductance-energy characteristics of a
diamond network with $15$ plaquettes for different values of $t_{so}$ in 
the absence of magnetic flux $\phi$. The $1$st, $2$nd and $3$rd rows
correspond to the results when $t_{so}=0$, $2$ and $4$, respectively.
The spin conserved conductances ($g_{\uparrow \uparrow}$) are plotted
in the first column, while in the second column spin flip conductances
($g_{\uparrow \downarrow}$) are given. In absence of $t_{so}$, gapless 
spectrum is observed for spin conserved conductance, while spin flip
conductance vanishes for the entire energy range. For all other cases,
a gap appears in the spectrum and its width can be regulated by tuning 
the Rashba coupling strength. 

The most interesting features in the conductance-energy characteristics 
are observed when we consider the effects of both the AB flux $\phi$ and 
Rashba SO coupling $t_{so}$. The results are shown in Fig.~\ref{cond4} for 
a diamond chain with $15$ identical diamond plaquettes for $\phi=0.4$ and
$t_{so}=4$. For sufficiently high AB flux and Rashba strength, two 
additional energy gaps occur at the flanks on both sides of the conductance 
spectrum in addition to the central gap. The energy gaps are positioned 
identically in all these spectra.

It is important to note that when anyone of $\phi$ and $t_{so}$ is zero 
and other is non-zero, $g_{\uparrow \uparrow}$ becomes exactly identical 
to $g_{\downarrow \downarrow}$, and so is $g_{\uparrow \downarrow}$ and 
$g_{\downarrow \uparrow}$. On the other hand, when both are non-zero,
spin conserved conductances differ in magnitude, but the spin flip
conductances remain identical. All these conductance-energy characteristics
shown in Figs.~\ref{cond1}-\ref{cond4} are compatible with the $E$-$k$
diagrams presented in Fig.~\ref{dispersion}. The gaps of the conductance
spectra of finite sized diamond chain compare well with those of the 
dispersion curves obtained earlier for an infinite sized diamond chain. 

\subsection{DOS-energy characteristics}

To gain insight into the nature of energy eigenstates of such a quantum 
network we address the behavior of average density of states. It is 
expressed as, 
\begin{equation}
\rho_{av}(E)=-\frac{1}{\pi N} {\mbox{Im}} [{\mbox{Tr}}[{\bf G}]]
\end{equation}
where, $N$ being the total number of atomic sites in the diamond chain. 

\begin{figure}[ht]
{\centering \resizebox*{9cm}{13cm}{\includegraphics{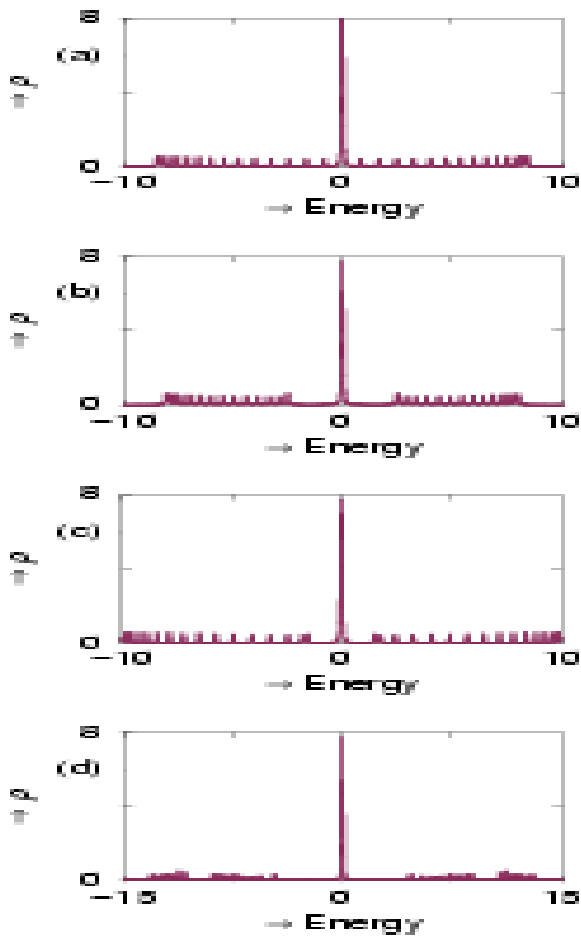}}\par}
\caption{(Color online). Average density of states as a function of energy 
$E$ for a diamond chain considering $15$ plaquettes with different values 
of $\phi$ and $t_{so}$ when $\epsilon_A=\epsilon_B=0$. (a) $\phi=0$, 
$t_{so}=0$; (b) $\phi=0.2$, $t_{so}=0$; (c) $\phi=0$, $t_{so}=2$ and 
(d) $\phi=0.4$, $t_{so}=4$.}
\label{dos1}
\end{figure}

As illustrative examples, in Fig.~\ref{dos1} we present the variations of
average density of states as a function of energy $E$ for a diamond chain 
consisting of $15$ plaquettes for different values of AB flux $\phi$ and 
Rashba SO coupling strength $t_{so}$. In (a), $\rho$-$E$ spectrum is given 
when both $\phi$ and $t_{so}$ are fixed at zero. The spectrum does not 
exhibit any gap as expected when $\epsilon_A=\epsilon_B=0$. A sharp peak 
is observed at the band center, i.e., at $E=0$ due to localized states. 
These localized states 
are highly degenerate and in general pinned at the energy $E=\epsilon_B$. 
The existence of the localized state is a characteristic feature of diamond
network as mentioned in an earlier work~\cite{sil}. In (b) and (c), 
energy gaps appear symmetrically around the central peak at $E=0$ by the
AB flux $\phi$ and Rashba coupling strength $t_{so}$, respectively. By 
tuning the AB flux $\phi$ or Rashba coupling strength $t_{so}$, the width of 
the gap can be controlled. Finally, in (d) we display average DOS when 
both $\phi$ and $t_{so}$ are finite. In such situation two extra gaps 
appear together with the central ones. The localized states are still 
situated at the same place as earlier.

\subsection{Effect of Rashba spin-orbit interaction on localization}

In such a quantum network, AB flux can induce complete localization. At 
$\phi=\phi_0/2$, conductance drops exactly to zero in the absence of 
Rashba SO interaction. This is due to the complete  
\begin{figure}[ht]
{\centering \resizebox*{8cm}{7cm}{\includegraphics{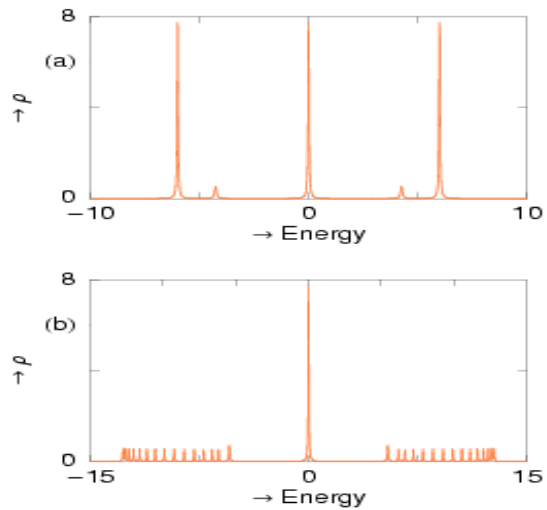}}\par}
\caption{(Color online). Average density of states as a function of 
energy for a diamond network considering $15$ plaquettes with 
$\epsilon_A=\epsilon_B=0$ when AB flux $\phi$ is set at $\phi_0/2$. 
(a) $t_{so}=0$ and (b) $t_{so}=2$.} 
\label{dos2}
\end{figure}
destructive interference between the electronic waves passing through
different arms of the network. At $\phi=\phi_0/2$, two more sharp peaks 
appear in the $\rho$-$E$ characteristics due to localized states 
(Fig.~\ref{dos2}(a)) in addition to the previous one pinned at $E=0$. 
The positions of these localized states can be evaluated exactly and 
they are expressed mathematically as,
\begin{equation}
E=\frac{1}{2} \left[(\epsilon_A + \epsilon_B) \pm \sqrt{(\epsilon_A + 
\epsilon_B)^2 - 4(\epsilon_A \epsilon_B - 4 t^2)}\right].
\label{equ26}
\end{equation}
In the presence of Rashba spin orbit interaction, the interference is 
not completely destructive anymore at $\phi=\phi_0/2$. The two additional
peaks at the opposite sides of the central one disappear for non-zero 
Rashba strength as clearly seen from Fig.~\ref{dos2}(b). Rashba spin-orbit 
coupling affects the spin dynamics significantly resulting in a non-zero 
conductance at $\phi=\phi_0/2$.

\subsection{Rashba induced semi-conducting behavior}

Here we address how Rashba SO interaction can induce semi-conducting
behavior in such a quantum network in the absence of $\phi$. A similar 
type of semi-conducting nature controlled by AB flux has been established 
in such a system, where SO interaction was not considered~\cite{sil}. To 
establish our idea, in
\begin{figure}[ht]
{\centering \resizebox*{8cm}{7cm}{\includegraphics{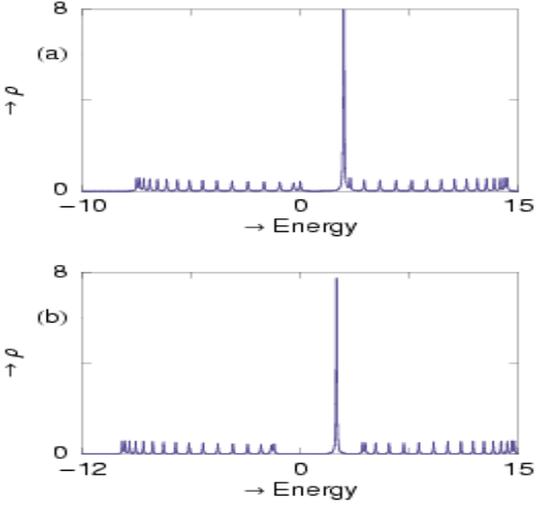}}\par}
\caption{(Color online). Average density of states as a function of energy
for a diamond chain considering $15$ plaquettes in the absence of AB 
flux $\phi$ where $\epsilon_A$ and $\epsilon_B$ are fixed at $0$ and $2$,
respectively. (a) $t_{so}=0$ and (b) $t_{so}=2$.}
\label{dos3}
\end{figure} 
Fig.~\ref{dos3} we plot the average density of states as a function of 
energy considering $15$ plaquettes where $\epsilon_A$ and $\epsilon_B$ 
are set at $0$ and $2$, respectively. When $\epsilon_A$ and $\epsilon_B$
are not same, the diamond network possesses an intrinsic gap in the
energy spectrum even in the absence of $\phi$ and $t_{so}$, as evident
from Fig.~\ref{dos3}(a). The sharp peak in the DOS is situated at the
edge of the gap, $E=\epsilon_B=2$ and it actually corresponds to the
localized states. By controlling the external 
gate potential, i.e., tuning the Rashba strength to a non-zero value, the 
width of the gap can be increased arbitrarily as shown in Fig.~\ref{dos3}(b),
keeping the position of the localized states invariant. Now, if the Fermi 
level $E_F$ is fixed at $E=2$ where we have localized states
(see Fig.~\ref{dos3}(b)), then for small Rashba coupling strength the gap 
between the localized level and the bottom of the right sub-band can be 
made small enough for the electrons to bridge. Therefore, the system behaves 
as a $n$-type semiconductor. Similarly, if $\epsilon_B$ is fixed at $-2$ 
and the Fermi level is set at the top of the left sub-band, then the system 
can be implemented equivalently as a $p$-type semiconductor. In this case
holes are created in the left sub-band. It is important to mention that 
when the site energies ($\epsilon_A$ and $\epsilon_B$) are fixed at the 
same value, then also the system can be used as a semi-conductor depending 
on the electron concentration. The detailed analysis is available 
in Ref.~\cite{sil}.

\subsection{Spin filtering action}

With proper tuning of the external parameters like, magnetic flux $\phi$ 
and Rashba strength $t_{so}$, a diamond network can achieve a high degree of
spin polarization as discussed earlier in a theoretical work by Aharony 
{\em et al.}~\cite{aharony}. Here, we discuss this feature from a 
different point of view. 

When there is no external magnetic field or magnetic flux, time reversal 
symmetry is not broken, and the Hamiltonian of the system remains 
\begin{figure}[ht]
{\centering \resizebox*{8cm}{7cm}{\includegraphics{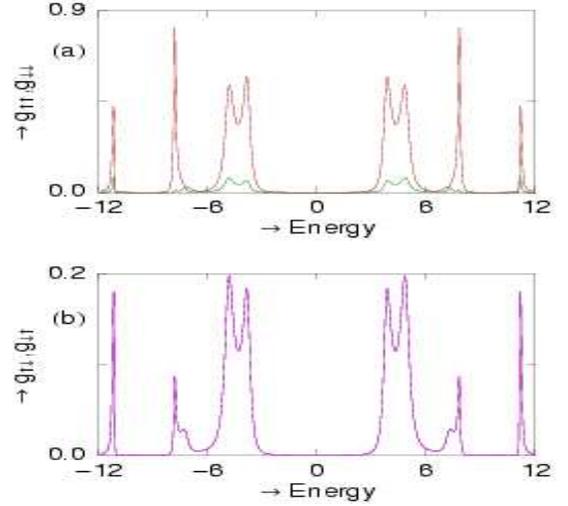}}\par}
\caption{(Color online). Variation of conductances as a function of energy
for a diamond chain with $3$ plaquettes considering $\phi=0.3$ and $t_{so}=4$
where $\epsilon_A=\epsilon_B=0$. 
In (a) $g_{\uparrow \uparrow}$ and $g_{\downarrow \downarrow}$ and in (b) 
$g_{\uparrow \downarrow}$ and $g_{\downarrow \uparrow}$ are superposed to 
each other.}
\label{filter}
\end{figure}
invariant under time reversal operation. Mathematically it is expressed 
as $[H_{SO},T]=0$, where $T$ is the time reversal operator.
The Rashba Hamiltonian ($H_{SO}$) is usually written as, 
\begin{eqnarray}
H_{SO} & = & \frac{\alpha_R}{\hbar} \left(\vec{\sigma} \times \vec{p}
\right)_Z \nonumber \\
& = & i \alpha_R \left( \sigma_y \frac{\partial}{\partial x} - \sigma_x 
\frac{\partial}{\partial y} \right) 
\label{equ33} 
\end{eqnarray}
and $T = i \sigma_{y} \hat {C}$, $\hat{C}$ being the complex conjugation 
operator. The second quantized form of Eq.~(\ref{equ33}) is given by the
fourth and fifth terms of Eq.~(\ref{equ2}). As electrons are 
spin-$\frac{1}{2}$ particles, so following 
Kramer's theorem, each eigenstate is at least two-fold degenerate and spin
is no longer a good quantum number in the presence of spin-orbit interaction.
{\em In many cases the degeneracy implied by Kramer's theorem is merely the
degeneracy between states of spin up and spin down, or something equally
obvious. The theorem is non-trivial for a system with spin-orbit coupling 
in an unsymmetrical electric field, so that neither nor angular momentum 
is conserved. Kramer's theorem implies that no such field can split the
degenerate pairs of energy levels}~\cite{ballen}.

However, the degeneracy can be removed by applying external magnetic flux
or magnetic field as in this case time reversal symmetry is not conserved
anymore and therefore spin polarization can be achieved. The degree of 
polarization of the transmitted electrons is conventionally defined as,
\begin{equation}
P(E)=\left| \frac{(g_{\uparrow \uparrow} + g_{\downarrow \uparrow})-
(g_{\downarrow \downarrow} + g_{\uparrow \downarrow})}{(g_{\uparrow 
\uparrow} + g_{\downarrow \uparrow})+(g_{\downarrow \downarrow} + 
g_{\uparrow \downarrow})} \right|.
\label{equ27}
\end{equation}

For our illustrative purpose, in Fig.~\ref{filter} we plot the variations
of conductances as a function of energy for a diamond network with three
plaquettes considering $\phi=0.3$ and $t_{so}=4$.
A significant change is observed in the magnitudes of spin conserved 
conductances ($g_{\uparrow \uparrow}$ and $g_{\downarrow \downarrow}$)
(Fig.~\ref{filter}(a)), while the spin flip conductances ($g_{\uparrow 
\downarrow}$ and $g_{\downarrow \uparrow}$) are identical as shown in 
Fig.~\ref{filter}(b). Therefore, applying a non-zero flux spin
polarization is clearly obtained. Following Eq.~(\ref{equ27}) we 
calculate the degree of polarization for an arbitrary energy $E=-5$ 
(say), and it is about $44\,\%$.

\section{Closing remarks}

To conclude, in the present work we have explored spin dependent
transport through an array of diamonds where Rashba SO interaction 
is present and each diamond plaquette is threaded by an AB flux $\phi$. 
The diamond chain is directly coupled to two semi-infinite $1$D non-magnetic 
metallic leads, namely, source and drain. We have adopted a discrete 
lattice model within the tight-binding framework to describe the system and 
present calculations based on Green's function formalism. We have obtained 
analytical expression for the $E$-$k$ dispersion relation for an infinite 
diamond network with Rashba SO interaction, and, show explicitly the 
interplay of spin-orbit interaction and magnetic flux on its band structure.
This analysis also gives insight about the presence of spin dependent 
localized and extended eigenstates which crucially controls the spin 
dependent transport through such device. This analytical study, in fact,
provides us a very good understanding about the transport behavior of
spins across a finite sized array of diamonds. It has been clearly 
established that how delocalizing effect sets in due to Rashba SO 
interaction when the AB flux $\phi$ is $\phi_0/2$. Quite interestingly
we show that depending on the specific choices of SO interaction strength 
and AB flux, the quantum network can be utilized as a spin filter. 

In the present work we have ignored the effects of temperature, 
electron-electron correlation, electron-phonon interaction, disorder, etc. 
Here, we set the temperature at $0$K, but the basic features will not 
change significantly even at low temperature as long as thermal energy 
($k_BT$) is less than the average level spacing of the diamond chain. 
In this model it is also assumed that the two side-attached non-magnetic 
leads have negligible resistance. Our presented results may be useful in 
designing spin based nano electronic devices.

\end{document}